\documentclass[twocolumn,showpacs,%
nofootinbib,aps,superscriptaddress,%
eqsecnum,prd,notitlepage,showkeys,10pt]{revtex4-1}

\usepackage{caption}
\usepackage{amssymb}
\usepackage{xcolor}
\usepackage{amsmath}
\usepackage{graphicx, wrapfig, lipsum, caption}
\usepackage{subfigure}
\usepackage{dcolumn}
\usepackage{hyperref}
\usepackage{blindtext}
\usepackage[sort&compress]{natbib}
\usepackage{setspace}
\usepackage{array}
\usepackage{booktabs}

\begin{document}

\title{Variable Goal Approach (VGA) Incorporating Human Intelligence into the Microscopic Pedestrian Dynamics Models}

\author{Kanika Jain$^1$, Anurag Tripathi$^1$, Shankar Prawesh$^2$, Indranil Saha Dalal}

\affiliation{Department of Chemical Engineering, Indian Institute of Technology Kanpur, India\\$^2$Department of Management Sciences, Indian Institute of Technology Kanpur, India}

\begin{abstract}
\textbf{Abstract:} Pedestrian dynamics models have provided valuable insights into pedestrian interactions, collision avoidance, and self-organized crowd behavior using mathematical, computational, AI-based, and heuristic approaches. However, existing models often fail to capture fundamental aspects of human decision-making, particularly the tendency to adopt indirect routes by sequentially selecting intermediate goals within the line of sight. In this study, we propose a novel Variable Goal Approach (VGA) that integrates human intelligence into pedestrian dynamics models by introducing multiple intermediate goals, termed variable goals, which guide pedestrians toward their final destination. These variable goals function as an adaptive guidance mechanism, enabling smoother transitions and dynamic navigation. VGA also enhances the efficiency of a model while minimizing interactions and disruptions. By strategically positioning variable goals, VGA introduces an element of stochasticity. This allows the model to simulate varied pedestrian paths under identical conditions, reflecting the diversity in human decision-making. In addition to its effectiveness in simple scenarios, VGA demonstrates strong performance in replicating high-density scenarios, such as lane formation, providing results that closely match real-world data.
\end{abstract}

\maketitle

\section{Introduction} \label{sec:Intro}

In the last several decades, pedestrian dynamics has garnered a growing amount of interest due to rising populations and substantial rural-to-urban migration. Managing these enormous crowds, particularly in mass meetings and public areas, is an incredibly difficult task that requires a design to ensure smooth operation and minimize risks in any untoward event. A further reason for the appeal of pedestrian dynamics is the expanding role of virtual reality in human life, such as Metaverse. The day in which robots are an integral part of our everyday life is not too distant. Sophia, a humanoid, is the finest illustration of this. In order to develop a more convenient and pleasant walking environment, a sound modeling of pedestrian flow is necessary.

Pedestrian dynamics modeling involves the simulation and analysis of pedestrian movements in various environments using mathematical \cite{helbing1995social, chraibi2010generalized, karamouzas2014universal}, computational \cite{burstedde2001simulation, van2011reciprocal, moussaid2011simple}, and AI-based \cite{alahi2016social, zhang2019sr} models. This field provides insights into pedestrian interactions, collision avoidance, and self-organization in crowded spaces. Various models have been proposed, differing in complexity, underlying assumptions, and applications. For instance, force-based models represent pedestrians as inertial particles subject to Newtonian-like forces (e.g., attraction, repulsion) to simulate complex scenarios such as self-organization \cite{helbing2001self}, panic behavior \cite{helbing2000simulating}, and evacuations \cite{helbing2002simulation}. Cellular automata models employ floor field concepts (static and dynamic), discretizing space into a grid where pedestrians transition between cells \cite{burstedde2001simulation, kirchner2002simulation, li2019review}. The Optimal Reciprocal Collision Avoidance (ORCA) model \cite{van2011reciprocal}, based on the velocity obstacle approach, ensures collision-free motion by selecting velocities that avoid imminent collision fields. This method is primarily applied in robotics and multi-agent systems \cite{cheng2017decentralized, dai2025mobile}. In recent years, deep learning techniques have gained prominence in pedestrian dynamics research \cite{alahi2016social, zhang2019sr, xue2018ss}. For example, the Social LSTM model leverages recurrent neural networks (RNNs) to learn social interactions directly from data \cite{alahi2016social}. Despite these sophisticated models, simpler rule-based approaches grounded in cognitive science also exist \cite{moussaid2011simple}. These models rely on behavioral heuristics, where pedestrians, guided by visual information, select a direction and velocity that facilitate the most direct path to their destination while maintaining a minimum time-to-collision threshold from obstacles. Such heuristics have also been applied to enhance autonomous robot navigation \cite{camara2020pedestrian, dorigo2021swarm}.

\begin{figure}
    \includegraphics[width= 1\linewidth]{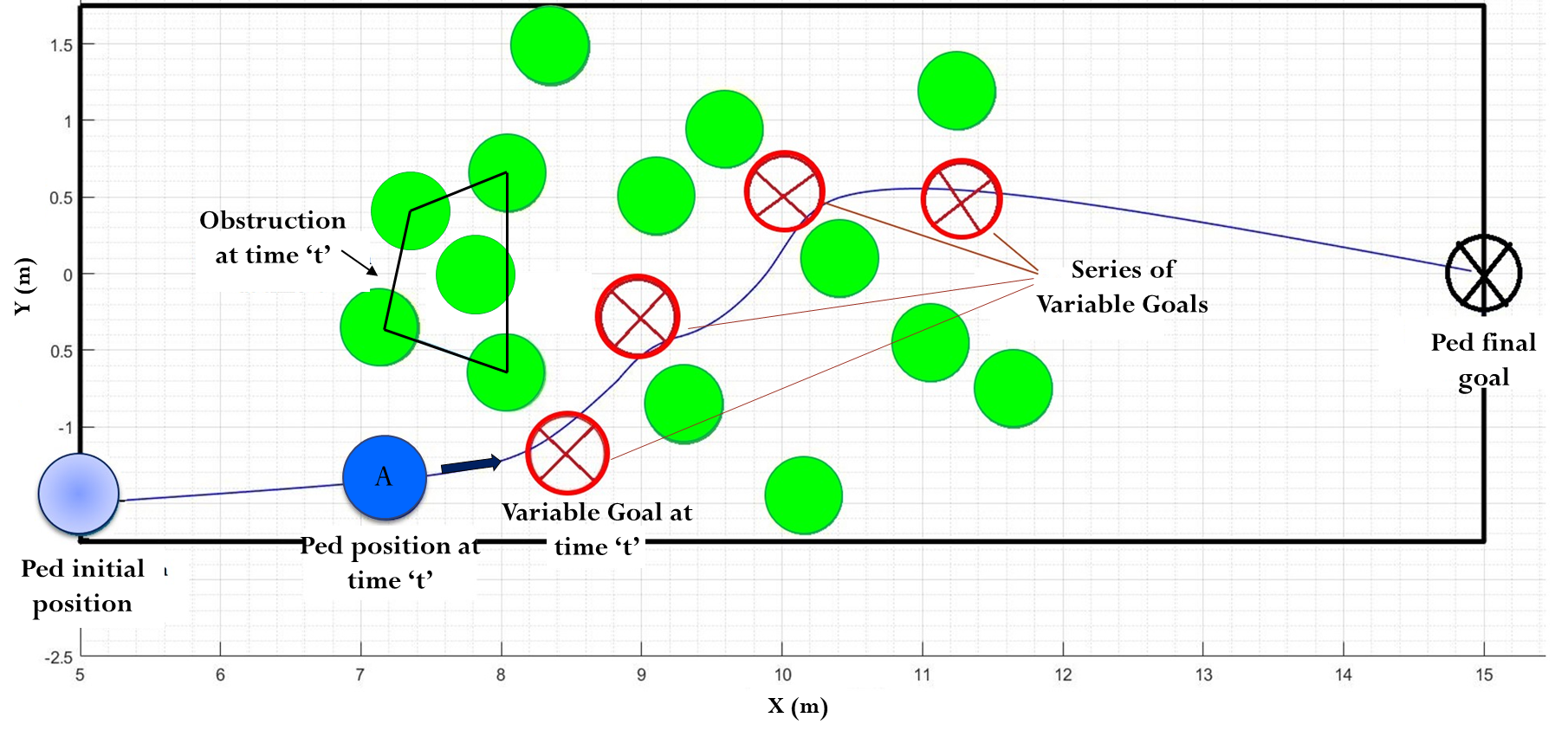}
    \caption{{\footnotesize Snapshot of the simulation demonstrating Variable Goal Approach (VGA) for the case of multiple obstacle single pedestrian (MOSP)}} \label{fig:VGA_snapshot}
\end{figure}

\begin{figure*}
    \includegraphics[width= 1\linewidth]{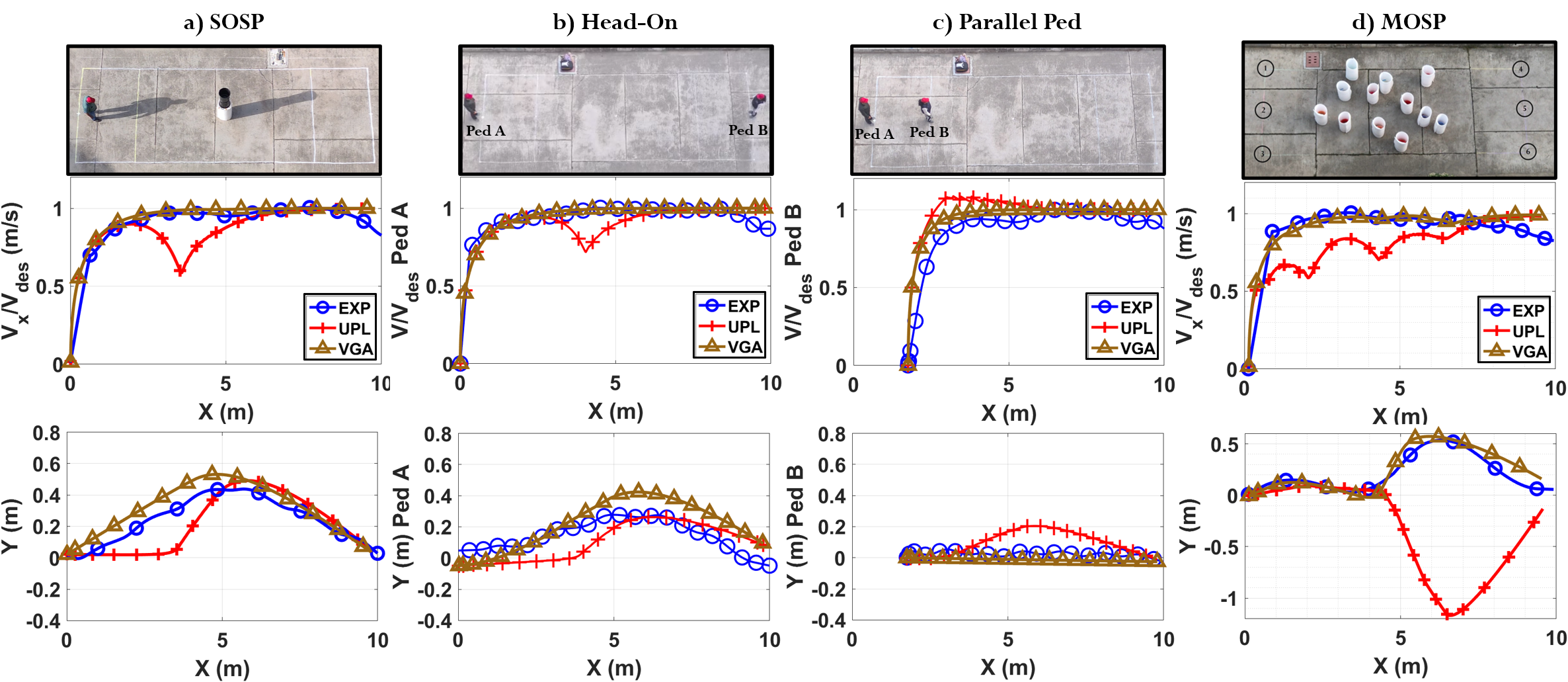}
    \caption{{\footnotesize Controlled experiments illustrating four real-life scenarios: (a) SOSP: A volunteer passes a stationary obstacle to reach a goal. (b) Head-On: Two volunteers swap positions while walking from opposite directions. (c) Parallel Ped: A faster volunteer overtakes a slower one. (d) MOSP: A volunteer navigates a maze of obstacles, with initial and final goals permuted among positions 1-6. Normalized speeds and trajectories are plotted for each scenario. UPL (red/plus) shows dips in speed and trajectory shifts in (a), (b), and (d), and a "push effect" in (c). After including VGA (golden/triangle), the model closely matches experimental data (blue/circle), avoiding sudden changes in speed and trajectory and other unrealistic behavior. Measurement areas range from 2 to 8 meters to avoid entry/exit effects.}} \label{fig:vxVSx}
\end{figure*}

Despite their wide applicability, existing pedestrian dynamics models often fail to incorporate fundamental aspects of human decision-making, particularly the tendency to adopt an indirect approach to reach the final destination. A pedestrian has a tendency to divide its path into multiple sections with the help of intermediate goals which are in direct line of sight. To address the same, we propose a novel Variable Goal Approach (VGA) which integrates ``human intelligence" into pedestrian dynamics modeling where the intermediate goals are termed as variable goals that enable natural path adjustments. Figure \ref{fig:VGA_snapshot} presents a schematic representation of a pedestrian navigating through a maze of obstacles, where variable goals facilitate collision avoidance and ensure a smooth transition toward the final destination. By leveraging variable goals, pedestrians can alter their direction without significant changes in fundamental behaviors such as speed. For instance, Figure \ref{fig:vxVSx} illustrates velocity profiles normalized by desired velocity (middle row) and corresponding trajectories (bottom row) across four basic real-life scenarios. The force-based model UPL exhibits unrealistic velocity fluctuations and abrupt trajectory deviations, whereas experimental results demonstrate stable velocities and smoother paths. The experimental data, obtained from our previous study \cite{jain2025benchmarking}, is publicly available at \url{https://github.com/kanika201293/Pedestrian-Experimental-Data}. These experimental results indicate that pedestrians anticipate collisions and adjust their paths in advance to ensure smooth transitions with minimal deviations. In contrast, force-based models rely on repulsive interaction forces to prevent collisions, often leading to unrealistic velocity changes and abrupt directional shifts, as shown in Figure \ref{fig:vxVSx}. In the VGA framework, pedestrians alter their direction by adjusting their goal rather than experiencing a large repulsive force due to obstacles. This mechanism allows for smoother and more natural motion, as evident in the graphs presented in Figure \ref{fig:vxVSx}.

The Variable Goal Approach (VGA) integrates human intelligence into pedestrian dynamics models by introducing the concept of variable goals. This approach enables a pedestrian

a) to divide their path into multiple segments before reaching the final destination,

b) to make decisions based on the visible surroundings,

c) to dynamically select goals, allowing flexible path direction adjustments,

d) to change direction without significantly altering fundamental behavioral characteristic, i.e., speed.

The article begins with the definition, conceptual framework, and stepwise implementation of the variable goal approach in Section \ref{sec:VGA}. Section \ref{sec:Results} presents various results obtained using VGA and discusses its ability to incorporate human intelligence into the model, enhance efficiency, introduce stochasticity, and replicate high-density scenarios. Finally, Section \ref{sec:Conclusion} summarizes the key findings and their implications.

\section{Variable Goal Approach (VGA)} \label{sec:VGA}

The variable goal approach is based on the principle of intermediate goals, termed as variable goals, which vary their positions with respect to the positions of the pedestrian and the obstruction.

\begin{figure}
    
    \centering
    
    \subfigure[Schematic diagram illustrating VGA for the case of Single Obstacle Single Pedestrian (SOSP).]
    {
        \includegraphics[width= 1\linewidth]{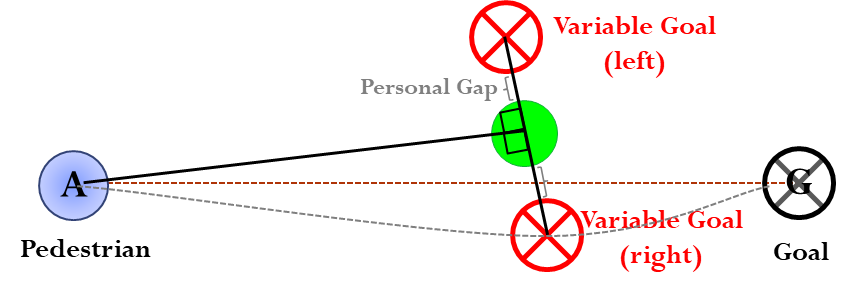}
        \label{fig:VGA_SOSP}
    }

    \subfigure[Schematic diagram depicting cluster formation and VGA for the case of multiple obstacles.]
    {
        \includegraphics[width= 1\linewidth]{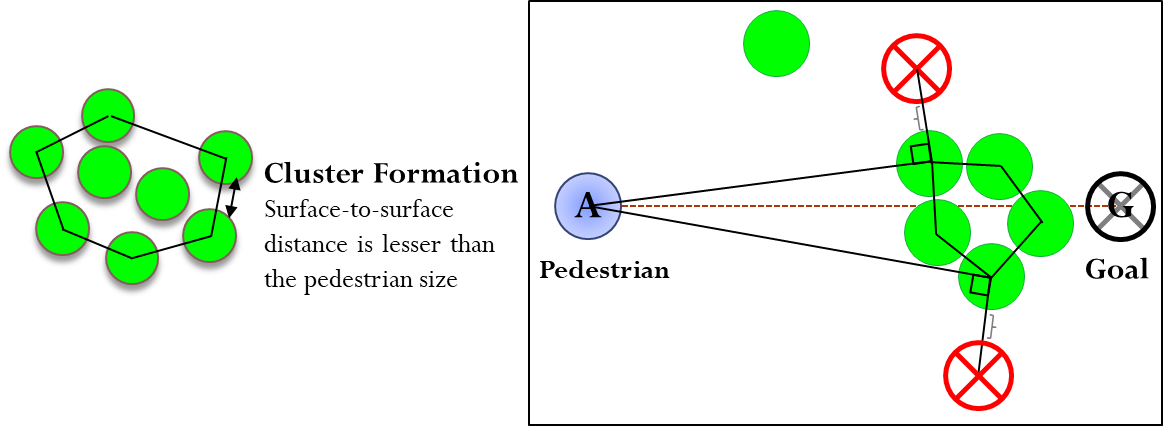}
        \label{fig:VGA_MOSP}
    }
    \caption{{\footnotesize 
    The figure explains the functioning of VGA in the case of a single obstacle as well as multiple obstacles. In the variable goal approach, two hypothetical goals ($\boldsymbol{\color{red} \otimes}$) are set perpendicularly on both sides of the obstacle or the cluster of obstacles maintaining a personal gap. (a) The figure shows VGA for the case of Single Obstacle Single Pedestrian (SOSP). The goal closest to the line joining the pedestrian `A' and the final goal `G' is selected using the least deviation method (here, variable goal (right). (b) When the surface-to-surface distance among the obstacles in lesser than the pedestrian size then the obstacles belong to one cluster. The obstacle cluster acts as one compact entity such that a pedestrian is unable to pass through. For multiple obstacles, the variable goals are set in the similar way and and the selection is done using the least deviation method.}}
    \label{fig:VGA_def}
\end{figure}

Consider a simple scenario of Single Obstacle Single Pedestrian (SOSP), where a single stationary obstacle obstructs the path of a single pedestrian `A' to reach its final goal `G', as depicted in Figure \ref{fig:VGA_SOSP}. To reach the final goal, pedestrian A must cross the obstacle either to the right or left. Now, according to VGA, a variable goal should be set to deal with the obstruction before reaching the final goal. To do so, two hypothetical goals are set perpendicularly on both sides of the obstacle with respect to the position of the pedestrian, as shown in Figure \ref{fig:VGA_SOSP} with symbol $\boldsymbol{\color{red} \otimes}$. The choice of one goal out of the two is made using the least deviation method. In this method, the goal, closer to the line joining the pedestrian and the final goal ($\overleftrightarrow {AG}$), is selected. In the figure, the right-hand side goal provides a lesser deviated path to the pedestrian. Thus, the right-hand side goal is chosen as the variable goal to reach the final destination with minimal interaction with obstacle.

For a more complex scenario involving multiple obstacles, a variable goal is set in the similar way. In this case, the two hypothetical goals are set perpendicularly on both sides of the cluster of obstacles, as shown in Figure \ref{fig:VGA_MOSP}. Obstacles are grouped into a cluster when their surface-to-surface distance is lesser than the pedestrian size. In this way, a cluster formation indicates a compact arrangement of obstacles such that a pedestrian is unable to pass through and the cluster can be treated as a single large obstacle. Here also, the variable goal is selected using the least deviation method to navigate around the cluster. 

\begin{figure}
    \includegraphics[width= 1\linewidth]{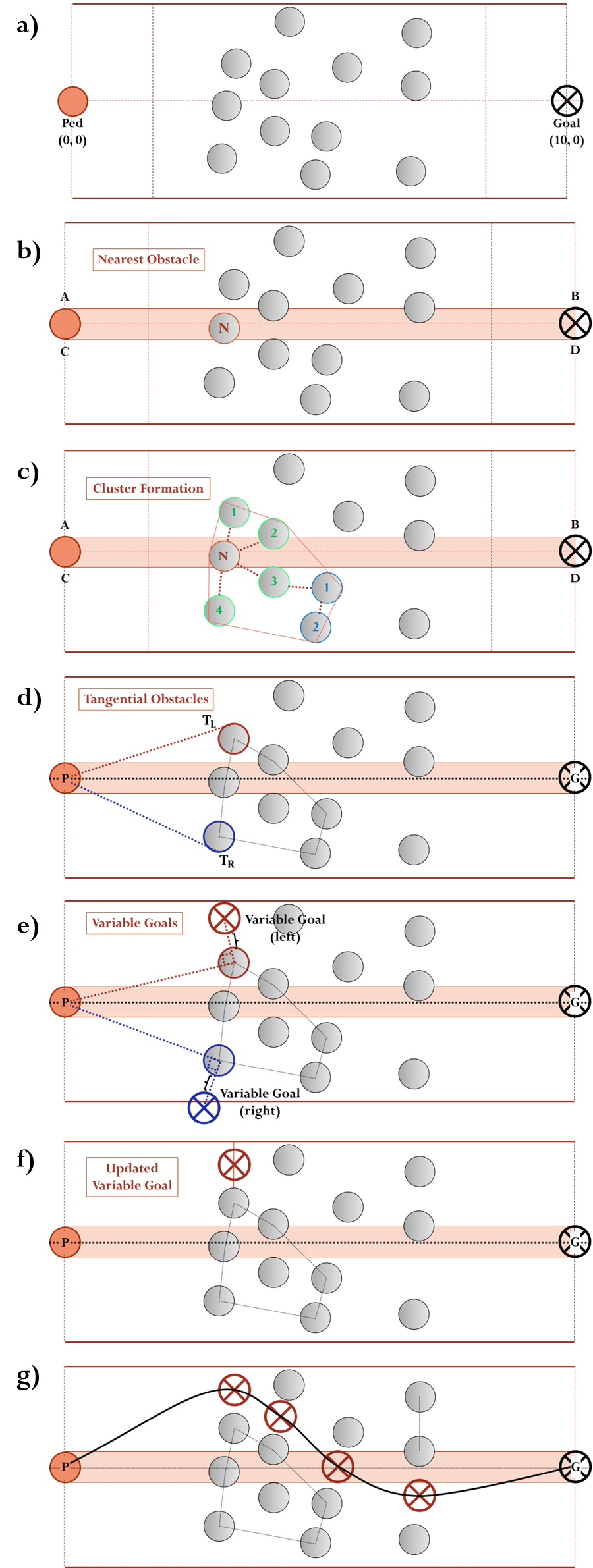}
    \caption{{\footnotesize Schematic diagrams explaining the step-wise implementation of variable goal approach for a smooth transition of a pedestrian.}} \label{fig:VGA_stepwise}
\end{figure}

A snapshot of a simulation in a multiple-obstacle single-pedestrian (MOSP) scenario is presented in Figure \ref{fig:VGA_snapshot}. As the pedestrian moves forward, it interacts with various obstructions, either individual obstacles or clusters of obstacles. The variable goals are set in a similar manner as mentioned earlier, allowing the pedestrian to make decisions up to the visible point and adjust its path direction while minimizing interactions with the obstructions. As shown in the figure, this approach enables the pedestrian to navigate smoothly through a maze of randomly placed obstacles, with the trajectory guided by a series of variable goals.\\\\
\textbf{Implementation:}\\
This subsection describes the process of determining the location of a variable goal to facilitate pedestrian navigation. The procedure consists of multiple steps: detection of the nearest obstacle, cluster formation, identification of tangential elements, determination of two possible variable goal locations, selection of one variable goal, and finally updating the variable goal position. Each step is elaborated below using the case of multiple obstacles, as illustrated in Figure \ref{fig:VGA_stepwise}a).\\\\
STEP 1: The process begins by identifying the nearest obstacle along the direct path between the pedestrian and the final goal. This is achieved by defining a hypothetical rectangular region, labeled A, B, C, and D in Figure \ref{fig:VGA_stepwise}b). Obstacles within this region are considered obstructions to the pedestrian’s path. Among these, the obstacle with the shortest center-to-center distance from the pedestrian is designated as the nearest obstacle (N).\\\\
STEP 2: Next, a cluster of obstacles surrounding the nearest obstacle (N) is formed. As previously defined, a cluster represents a compact arrangement of obstacles through which a pedestrian cannot pass. Obstacles whose surface-to-surface distance from N is smaller than the pedestrian size are included in the cluster. In Figure \ref{fig:VGA_stepwise}c), the four green-outlined obstacles represent this cluster. The surface-to-surface distances of these obstacles are then compared against other obstacles in the vicinity, and those within the pedestrian size threshold are also included to the cluster (e.g., two blue-lined obstacles). This iterative process continues until no additional obstacles can be added. As a result, the pedestrian perceives the cluster as a single, larger obstacle to be avoided.\\\\
STEP 3: The tangential obstacles of the cluster in both left and right directions are then identified, denoted as $T_L$ and $T_R$ in Figure \ref{fig:VGA_stepwise}d). The left tangential obstacle ($T_L$) is the obstacle forming the leftmost angle ($\angle T_LPG$) with respect to the pedestrian's position, while the right tangential obstacle ($T_R$) forms the rightmost angle ($\angle T_RPG$).\\\\
STEP 4: Once the tangential obstacles are determined, variable goals (represented as pedestrian-sized circles) are positioned perpendicularly to these obstacles. Figure \ref{fig:VGA_stepwise}e illustrates two such variable goals, placed at right angles to the centers of the tangential obstacles while maintaining a personal gap of half the pedestrian size \cite{}. If the tangential obstacles on both sides are the same, variable goals are positioned on both sides of the same obstacle, as observed in the case of SOSP in Figure \ref{fig:VGA_SOSP}.\\\\
STEP 5: Selecting one of the two variable goals is a critical step, requiring the following considerations:

a) The center of the variable goal must not extend beyond the corridor, ensuring sufficient navigable space.

b) The variable goal within the pedestrian's visible vicinity ($-100^{\circ}$ to $100^{\circ}$) is prioritized.

c) If both variable goals lie within the corridor and the visible vicinity, the selection is based on the least deviation method, whereby the goal closest to the line connecting the pedestrian and the final goal ($\overleftrightarrow {AG}$) is chosen.\\
Note that the proposed approach fails only when both variable goals are located outside the corridor. If a variable goal is outside the corridor and the other remains within the corridor but is outside the pedestrian’s visible vicinity, the goal outside the visible vicinity is selected. This selection indicates a stuck condition in the visible area, prompting the pedestrian to reorient and find a new path. For further clarity, refer to Figure S1. In Figure \ref{fig:VGA_stepwise}e), the variable goal on the right violates the corridor constraint; therefore, the variable goal on the left is selected.\\\\
STEP 6- After selection, the variable goal’s position is refined based on its surroundings. If it fails to maintain the personal gap with the surroundings also, such as other obstacles and corridors, then the position of the variable goal is adjusted to ensure equal spacing on both sides. For example, in Figure \ref{fig:VGA_stepwise}e), the selected variable goal is too close to the corridor. Consequently, its position is updated to the midpoint between the obstacle and the corridor, maintaining equal gaps on both sides (Figure \ref{fig:VGA_stepwise}f).\\

By iteratively repeating these six steps — ranging from identifying the nearest obstacle to updating the variable goal position — the locations of subsequent variable goals are determined. Figure \ref{fig:VGA_stepwise}g) illustrates the pedestrian's trajectory guided by a series of variable goals. This iterative process enables the Variable Goal Approach to effectively navigate pedestrians toward their final destination while minimizing obstructions.

\section{Results and discussion} \label{sec:Results}

VGA utilizes multiple variable goals to facilitate a smooth transition of a pedestrian from the initial to the final position. This section presents various results obtained by applying the variable goal approach to UPL, a force-based model, to highlight the significance of the approach. The results demonstrate four key characteristics of VGA, as discussed below.

\begin{figure}
    \includegraphics[width= 1\linewidth]{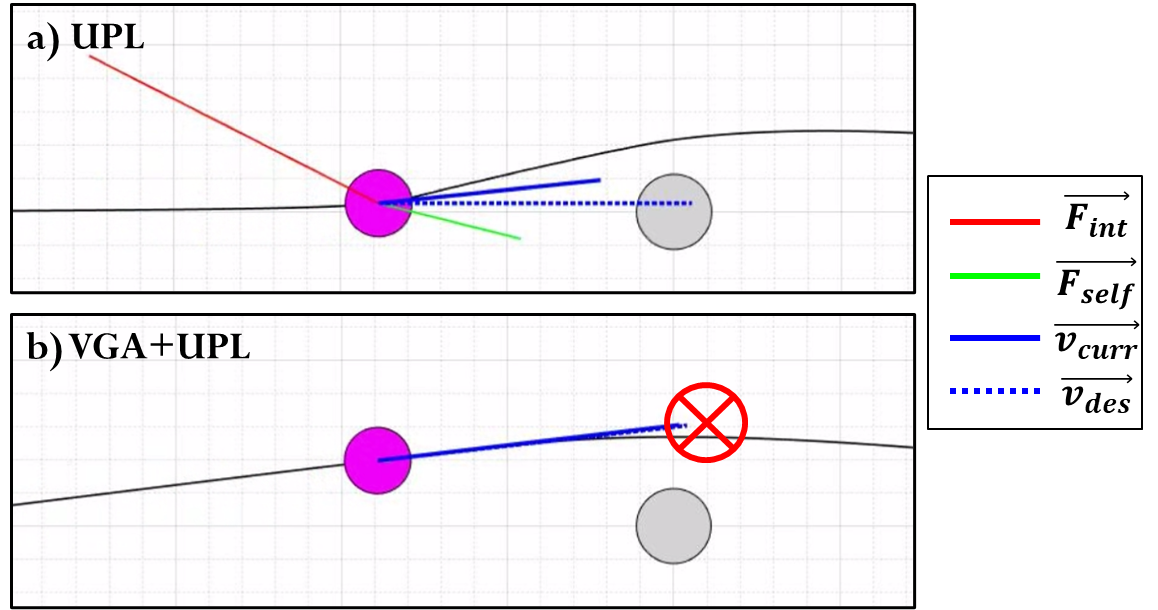}
    \caption{{\footnotesize Schematic diagram of forces experienced by a pedestrian in SOSP scenario while simulating using a) UPL alone and b) by introducing VGA to the model UPL. Here, $\overrightarrow{F_{int}}$ is interaction force, $\overrightarrow{F_{self}}$ is self-driving force, $\overrightarrow{v_{curr}}$ is direction of current velocity, and $\overrightarrow{v_{des}}$ is the direction of desired velocity. Note that the length of line segments is proportional to the respective magnitudes. The figure shows a drastic change in pedestrian experiencing forces due to interaction with an obstacle, by introducing VGA to the model.}} \label{fig:forces}
\end{figure}

\subsection{VGA: incorporating human intelligence}

As explained earlier, variable goals enable the incorporation of human intelligence into the model, allowing pedestrians to transition with minimal interaction with obstacles. Figure \ref{fig:forces} illustrates an SOSP scenario, comparing the interaction force experienced by a pedestrian due to an obstacle, in simulations with and without the variable goal approach. The figure shows that without VGA, the pedestrian encounters excessive repulsive interaction force while avoiding the obstacle, leading to a reduction in velocity and abrupt changes in path direction. In contrast, with VGA, the pedestrian successfully avoids the obstacle using a variable goal without experiencing interaction force. This avoids unrealistic changes in velocity or path direction. The length of the line segments in Figure \ref{fig:forces} is proportional to the magnitudes of the respective quantities. The scoring system developed in our previous study \cite{jain2025benchmarking} evaluated UPL with a score below 60\%, whereas after applying VGA, the score increased to over 90\%. This improvement demonstrates the incorporation of human intelligence into the model, enhancing different parameters such as oscillation, path smoothness, and speed deviation (see Table S1).

\subsection{VGA: increasing efficiency}

\begin{figure*}
    \includegraphics[width= 0.75\linewidth]{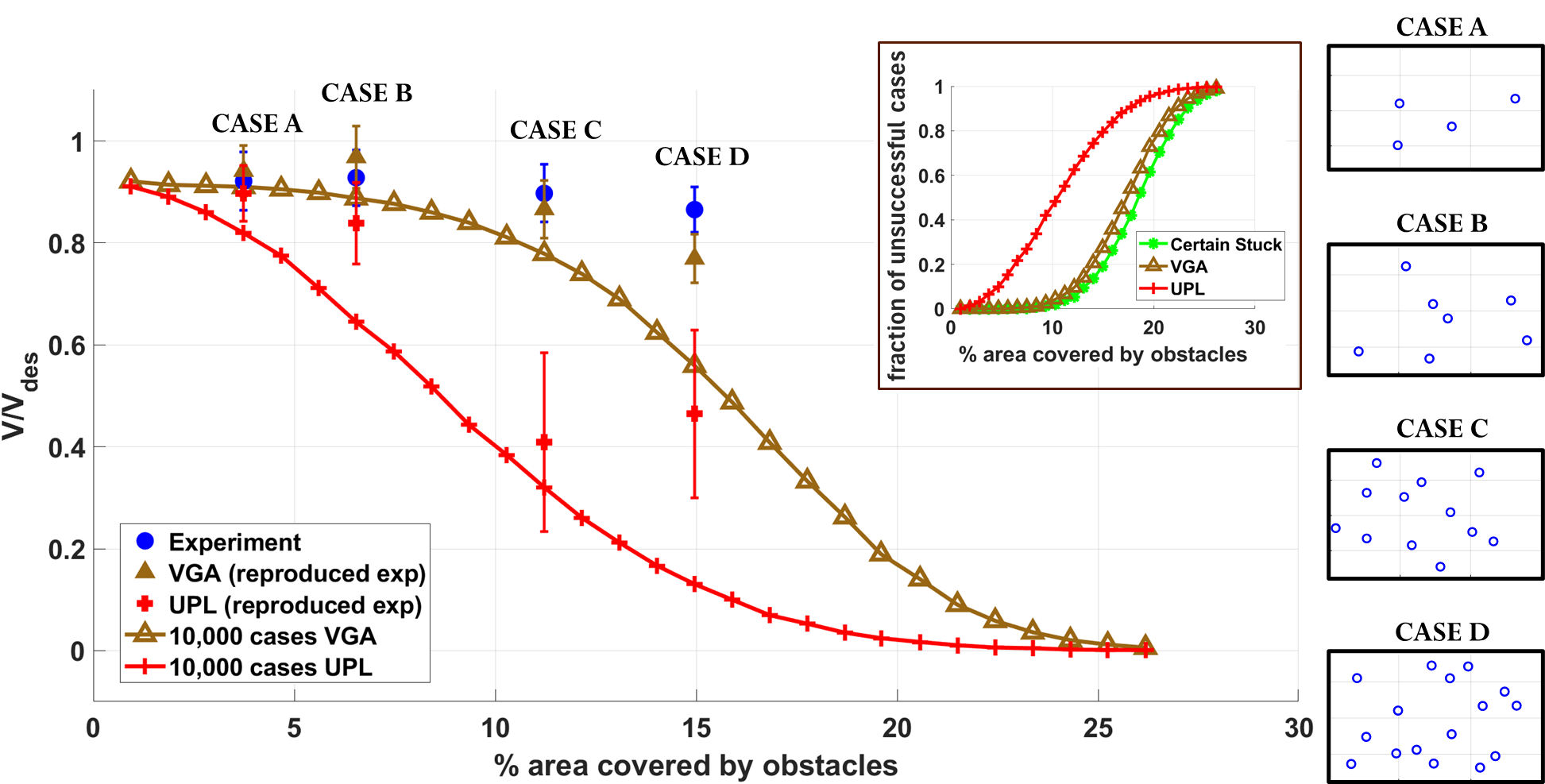}
    \caption{{\footnotesize The figure shows average normalized velocities with error bars for both experimental and simulated data across four obstacle conditions (Case A, B, C, D) in the MOSP experiment. Further, simulations are performed across obstacle area percentages (0\% to 27\%), with 10,000 configurations per percentage. The normalized velocities averaged over these 10,000 cases show that the model is able to maintain pedestrian speed more effectively with the help of VGA. The inset graph shows the fraction of unsuccessful cases out of 10,000. The green star-marked curve represents the fraction of cases having obstacle configurations with no possible path, indicating certain stuck conditions. The close alignment between the VGA curve and the certain stuck curve indicates that most failures are due to obstacle configuration rather than a limitation of the VGA.}} \label{fig:probabilityGraph}
\end{figure*}

\begin{figure*}
    \includegraphics[width= 0.75\linewidth]{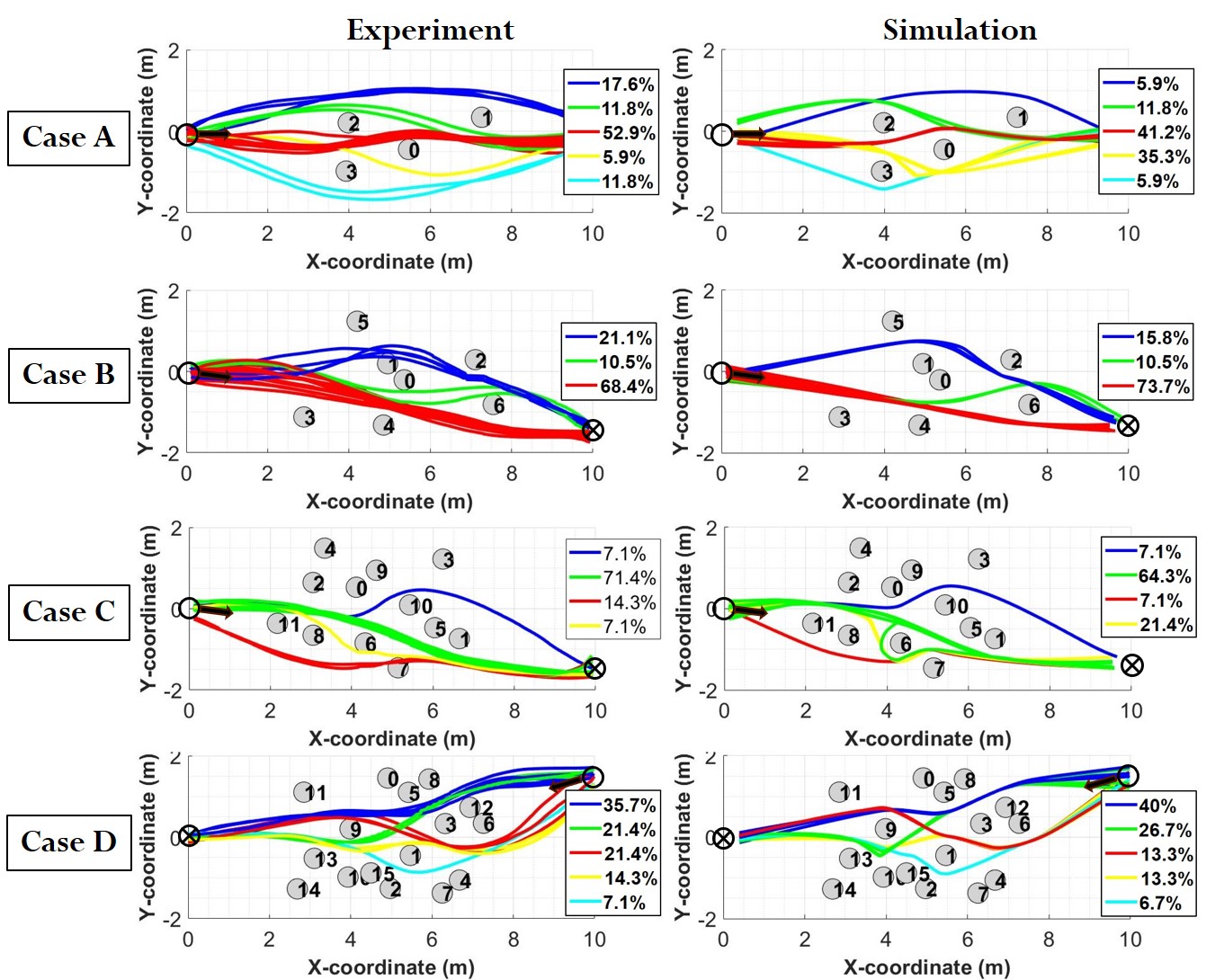}
    \caption{{\footnotesize Despite identical starting and ending positions, volunteers in the experiment followed different paths during the experiments (left). Each path is differentiated using different colors. Similar multiple paths are generated using VGA by introducing a probability mechanism during the selection of variable goal (right). The percentage of choosing a particular path in each scenario is specified.}} \label{fig:VGA_stochastic}
\end{figure*}

Beyond incorporating human intelligence, VGA significantly enhances model efficiency. In the MOSP scenario, where a pedestrian navigates through a randomly placed maze of obstacles, experiments were conducted with four different area percentages covered by obstacles using a specific obstacle configuration \cite{jain2025benchmarking}. These conditions, labeled Cases A, B, C, and D, are shown in Figure \ref{fig:probabilityGraph}. The figure presents average normalized velocities, along with error bars, for experimental data (blue circles), UPL-alone simulations (red crosses), and VGA+UPL simulations (gold triangles). The results indicate that VGA improves model performance by producing simulations that closely align with experimental data, within error limits, whereas UPL alone shows significant deviations, particularly at higher obstacle densities.

Since conducting experiments for every possible area percentage and obstacle configuration is impractical, simulations were extended to cover area percentages from 0\% to 27\%, with 10,000 different obstacle configurations for each percentage. The averaged normalized velocities across these configurations, plotted in Figure \ref{fig:probabilityGraph}, demonstrate that VGA enables the model to maintain pedestrian speed more effectively while avoiding obstacles compared to UPL alone.

The inset of Figure \ref{fig:probabilityGraph} plots the fraction of unsuccessful cases against the obstacle-covered area. The green star-marked curve represents cases where the obstacle configuration is in such a way that no feasible path exists, indicating inherently stuck conditions. The close agreement between this curve and the unsuccessful case fraction for VGA suggests that most failures in VGA simulations result from the obstacle configuration rather than a limitation of the approach. In contrast, UPL alone exhibits approximately 50\% failure at just 10\% obstacle coverage, indicating model inadequacy. Thus, VGA significantly improves the model’s efficiency.

\subsection{VGA: introducing stochasticity}

Pedestrians exhibit diverse behaviors, even when starting from the same location and heading toward the same destination. This variability was observed in MOSP experiments, where participants followed different paths despite identical initial and final positions. Figure \ref{fig:VGA_stochastic} (left) highlights these variations, illustrating the stochastic nature of pedestrian dynamics. A similar stochasticity can be incorporated into VGA through the selection of intermediate goals.

Previously, goal selection was based on the least deviation principle. However, by introducing a probabilistic bias as a function of deviation, a stochastic version of VGA can be obtained. Rather than strictly minimizing deviation, goal selection can follow a probabilistic distribution, allowing the model to generate multiple possible paths for the same start and end positions, as shown in Figure \ref{fig:VGA_stochastic} (right). If $d_L$ and $d_R$ represent deviations for the left and right goals, the selection probabilities are given by $P_L = d_R/(d_L +d_R)$ and $P_R = d_L/(d_L +d_R)$, making probability inversely proportional to deviation. Thus, if $d_L < d_R$, the left variable goal is more likely to be chosen.

A key challenge remains in determining the precise probability distribution for path selection to better replicate pedestrian decision-making in real-world scenarios. Further research, including additional data collection, is required to refine the goal selection process and enhance the model’s accuracy in capturing pedestrian behavior.

\subsection{VGA: replicating high-density scenarios}

\begin{figure}
    \includegraphics[width= 0.75\linewidth]{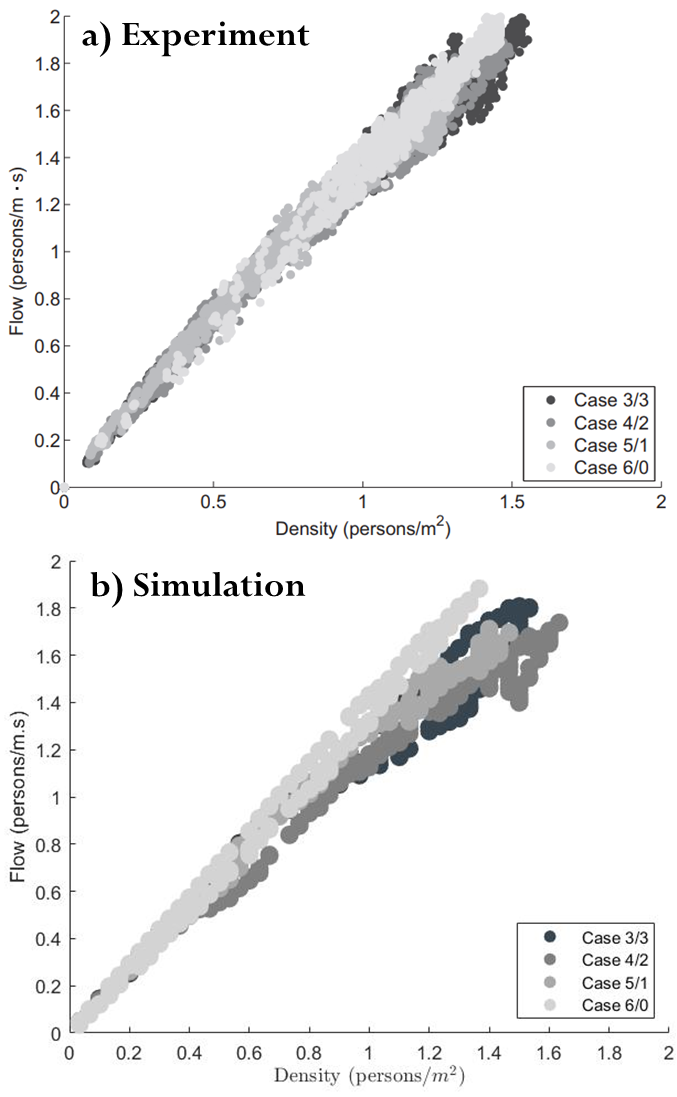}
    \caption{{\footnotesize Fundamental diagrams illustrating the relationship between the pedestrian flow and density is produced by (a) experimental data and (b) simulation using Variable Goal Approach.}} \label{fig:highDensity}
\end{figure}

While this study primarily focuses on low to moderate density scenarios, evaluating VGA’s performance in multi-pedestrian settings is essential. Here, we assess VGA in a bidirectional flow scenario. A previous experimental study on lane formation in opposing pedestrian streams at varying densities \cite{feliciani2016empirical} serves as a benchmark. The experiment is reproduced using VGA with UPL-based interaction forces, and a fundamental diagram illustrating pedestrian flow versus density is generated (see Figure \ref{fig:highDensity}). The results closely align with experimental observations, demonstrating VGA’s ability to model pedestrian interactions in high-density conditions.

To optimize computational efficiency in generating variable goals for each pedestrian at every time step, a maximum time-to-collision threshold of 3 seconds is applied \cite{karamouzas2014universal}. Additionally, visual snapshots of VGA-based simulations are presented in Figure S2, illustrating the emergence of multiple lanes at varying crowd stream densities. These results provide further empirical evidence of VGA’s capability to simulate emergent behaviors under high-density conditions. Overall, these findings highlight VGA’s potential as a robust and reliable tool for studying and modeling pedestrian dynamics in both single and multi-pedestrian scenarios.

\section{Conclusion} \label{sec:Conclusion}

The Variable Goal Approach (VGA) enhances microscopic pedestrian dynamics models by embedding human-intelligence using intermediate goals, providing a more realistic and efficient framework for pedestrian navigation. This approach allows pedestrians to alter their direction of motion by adding an intermediate goal resulting in minimal interaction with the obstruction, closely mirroring real-life pedestrian behavior. As mentioned in Section \ref{sec:VGA}, VGA introduces intermediate goals as variable goals, that act as a guidance system, facilitating smoother transitions toward the final goal while minimizing interactions.

VGA incorporates human intelligence into the model and significantly improves efficiency. By adding a probabilistic mechanism for the selection of the variable goals, VGA can introduce an element of stochasticity to the model, which enables the model to produce multiple pedestrian paths that reflect the diversity of human decision-making. In addition to the strong performance in fewer pedestrian scenarios, VGA is also capable of accurately replicating multiple pedestrian scenarios, closely aligning with reality, as elaborated in subsequent sections.

\bibliographystyle{ieeetr}
\bibliography{references}

\end{document}